\documentstyle[aps,prb,multicol,bezier]{revtex}

\begin{document}
\title{
   The probability for a
   wave packet to remain in a disordered cavity .
}

\author{ Daniel L. Miller }
\address{
 Dept. of Physics of Complex Systems,\\
 The Weizmann Institute of science,
 Rehovot, 76100 Israel                \\
 e-mail  fndaniil@wicc.weizmann.ac.il
}

\date{\today}

\maketitle
\begin{abstract}
   We show that the probability that a wave packet will remain in a
   disordered cavity until the time $t$ decreases exponentially for times shorter
   than the Heisenberg time and log-normally for times much longer than the
   Heisenberg time. Our result is equivalent to the known result for
   time-dependent conductance; in particular, it is independent of the
   dimensionality of the cavity. We perform non-perturbative ensemble
   averaging over disorder by making use of field theory. We make use of
   a one-mode approximation which also gives an interpolation
   formula (arccosh-normal distribution) for the probability to remain.
   We have checked that the optimal fluctuation method gives the same
   result for the particular geometry which we have chosen.
   We also show that the probability to remain does not relate simply to the
   form-factor of the delay time. Finally, we give an interpretation of the
   result in terms of path integrals. 
   \\
\end{abstract}

\begin{multicols}{2}
\narrowtext

The interest of experimentalists in open quantum dots\cite{Marcus-97}
motivates the computation of the probability to remain in a weakly
disordered cavity. This problem is similar to the computation of the time
dependent conductance of weakly disordered
samples\cite{AKL-feb87,MK-oct94,Mirlin-oct95}. However the geometry of our
problem allows a simple computation scheme, which is similar to the
zero-dimensional non-linear $\sigma$-model.

Let us assume that we have a weakly disordered open cavity and the whole
system is filled by non-interacting fermions at zero temperature.  At
time $t=0$ the Fermi level outside of the cavity decreases and particles
begin to escape from the cavity.  The probability for the particles to remain
in the cavity of volume $S$ till time $t$ is given by the ratio
\begin{equation}
   p(t) = {\int_S \varrho(\vec r, t) d\vec r \over
           \int_S \varrho(\vec r, 0) d\vec r }\;,
\label{eq:class.1}
\end{equation}
where $\rho(\vec r, t)$ is the density of the particles on the Fermi surface,
and $\rho(\vec r, 0)=1$ for $\vec r\in S$ and 0 elsewhere.
On a short
time scale this is the standard problem of the diffusion
emission.\cite{Morse-Feshbach-book} The time evolution of the distribution
function can be found by solving the diffusion equation
\begin{equation}
   \left[ {\partial \over \partial t} - D\nabla^2 \right] \varrho(\vec r, t)
   =0
\label{eq:class.2}
\end{equation}
with the boundary conditions $\vec n \vec \nabla \varrho =0$ at the walls of the
cavity ($\vec n$ is normal to the boundary) and
$\varrho + 0.71 \ell \vec n \vec \nabla \varrho = 0 $ at the open edge
of the cavity.\cite{Morse-Feshbach-book} Here  $\ell$ is the mean free path
and the scattering is uniform.
In further calculations we will use the open edge boundary condition
$\varrho=0$, assuming that $\ell$ is much smaller than all the characteristic
lengths of the system.

\begin{figure}
\unitlength=1.00mm
\linethickness{0.4pt}
\begin{picture}(65.00,60.00)
\put(40.00,30.00){\circle{10.00}}
\put(47.00,30.00){\makebox(0,0)[lc]{$\varrho=0$}}
\bezier{200}(15.00,30.00)(15.00,5.00)(40.00,5.00)
\bezier{200}(15.00,30.00)(15.00,55.00)(40.00,55.00)
\bezier{200}(40.00,55.00)(65.00,55.00)(65.00,30.00)
\bezier{200}(65.00,30.00)(65.00,5.00)(40.00,5.00)
\put(63.00,47.00){\makebox(0,0)[lb]{$\nabla_r\varrho=0$}}
\put(40.00,30.00){\vector(0,1){5.00}}
\put(40.00,37.00){\makebox(0,0)[cb]{$L_1$}}
\put(40.00,30.00){\vector(-1,1){19.00}}
\put(20.00,50.00){\makebox(0,0)[rb]{$L_2$}}
\end{picture}
\caption{Boundary conditions for the diffusion mode in the circular cavity
with the contact in the middle.}
\label{fig:circular}
\end{figure}
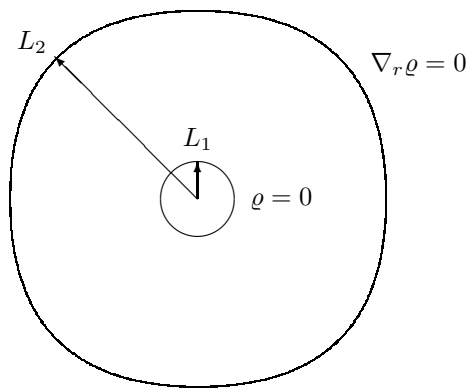

The solution of the diffusion equation can be represented as a sum over
diffusion modes, each of them decaying exponentially at its own rate. The
``lowest'' diffusion mode is computed in Appendix~\ref{sec:App.A} for two
examples of circular and spherical cavities with the contact in the middle,
see Fig.~\ref{fig:circular}. Therefore the probability to remain in the cavity
behaves like $p(t)= e^{-\gamma t}$, where $\gamma$ is the decay rate of the
``lowest'' diffusion mode. This escape rate is proportional to the
size of the contact and may vary from zero to the inverse diffusion time
through the system $E_c/\hbar$.  In the rest of the paper we will use 
units where $\hbar=1$, and then we have in general
\begin{equation}
    0 \le \gamma \le E_c\;.
\label{eq:class.3}
\end{equation}

In quantum mechanics the evolution of the density matrix is given by the
product of two exact Green functions
\begin{equation}
   \rho(\vec r_1,\vec r_4, t) =
   \int d\vec r_2d\vec r_3 G^R(\vec r_1,\vec r_2,t) G^A(\vec r_3,\vec r_4,t)
   \rho(\vec r_2,\vec r_3, 0)\;.
\label{eq:quant.1}
\end{equation}
In order to compute the probability to remain in the cavity 
we should prepare our system in
the state
\begin{eqnarray}
   \rho(\vec r_2,\vec r_3, 0) &=& \int d\vec p\,
   e^{i\vec p (\vec r_2 - \vec r_3)}\;
   \varrho({\vec r_2+\vec r_3 \over 2}, 0)\;,
\nonumber\\
   &&
   \delta(E_F - {\cal H}(\vec p,{\vec r_2+\vec r_3 \over 2}))
\label{eq:quant.2}
\end{eqnarray}
where ${\cal H}(\vec p,\vec r)$ is the Hamiltonian of the system, and
$\varrho(\vec r,0)$ is one inside the cavity and zero outside. The
quantum-mechanical expression for the probability to remain is therefore
\begin{eqnarray}
   p(t) &=& {\int_S \rho(\vec r, \vec r, t) d\vec r \over
           \int_S \rho(\vec r, \vec r, 0) d\vec r }
    = {1\over 2\pi \nu S } \int{d\omega\over 2\pi}\, e^{-i\omega t}\,
\nonumber\\ &\times&
      \int_S d\vec r_1 d\vec r_2
      G^R_{E_F+{\omega\over 2}}(\vec r_1, \vec r_2)
     G^A_{E_F-{\omega\over 2}}(\vec r_2, \vec r_1)\;,
\label{eq:quant.3}
\end{eqnarray}
where $\nu$ is the density of states on the Fermi surface.
The exact Green functions depend on the realizations of
the disorder potential. The non-perturbative averaging over the
$\delta$-correlated disorder potential can be done by making use of field
theory\cite{Efetov-book97}.  Our main result is
\begin{equation}
  p(t) = \exp\left\{-{\pi \gamma\over 2\Delta}
  \mbox{arccosh}^2\left({t\Delta\over\pi}+1\right)\right\}\;,
\label{eq:quant.4}
\end{equation}
where $\Delta$ is the mean level spacing of the ``closed'' cavity $1/\Delta =
\nu S$ and $\gamma$ is the classical escape
rate. This formula gives exponential decay on the short time scale and log --
normal decay on the long time scale. Equation~(\ref{eq:quant.4}) was derived
under the conditions
\begin{mathletters}
\begin{eqnarray}
   \gamma &\ll& E_c\;,
\label{eq:quant.4a}
\\
   t &\gg& E_c^{-1}\;,
\label{eq:quant.4b}
\\
    \log(t\Delta)&\ll& \left\{
    \begin{array}{cc}
       L_1\log(L_2/L_1)/\ell & \text{ 2 dimensions},\\
       L_1/\ell & \text{ 3 dimensions},
    \end{array}\right.
\label{eq:quant.4c}
\end{eqnarray}
for all relation of $\gamma$ and $\Delta$. Particularly, the exact
probability to remain should have some features on the time scale $E_c^{-1}$
near $t=0$ and $t=2\pi/\Delta$, as it takes place for spectral form-factors
of closed systems\cite{Andreev-Altshuler-95}. Our result
Eq.~(\ref{eq:quant.4}) does not have any structure on the time scale
$E_c^{-1}$ and this is the meaning of the conditions Eqs.~(\ref{eq:quant.4a})
and (\ref{eq:quant.4b}). If the last inequality Eq.~(\ref{eq:quant.4c}) is 
not fulfilled one should use
the ballistic action\cite{MK-jul95} in the field theory.
\end{mathletters}

It is interesting to compare the result Eq.~(\ref{eq:quant.4}) with the
form-factor of the delay time
\begin{eqnarray}
   K_\gamma(t) &=&
   {1\over 2\pi \nu S } \int{d\omega\over 2\pi}\, e^{-i\omega t}
   {\cal T}(E_F+{\omega\over 2})
   {\cal T}(E_F-{\omega\over 2})
\nonumber\\ &\approx&
   {1\over 2\pi \nu S } \int{d\omega\over 2\pi}\, e^{-i\omega t}\,
\nonumber\\ &\times&
      \int_S d\vec r_1 d\vec r_2
      G^R_{E_F+{\omega\over 2}}(\vec r_1, \vec r_1)
     G^A_{E_F-{\omega\over 2}}(\vec r_2, \vec r_2)\;,
\label{eq:form.1}
\end{eqnarray}
where the subscript $\gamma$ means that our system is open (it has the classical
escape rate $\gamma$). The Schroedinger equation for the cavity has solutions,
which are expanding waves far away from the system. The corresponding 
eigenvalues of the energy are complex $E_n-i\Gamma_n/2$, see
Ref.\onlinecite{Landau-QM}, \S134. The delay time is the analog of the density
of states ${\cal T}(E) = \text{Im}\sum_n (E-E_n+i\Gamma_n/2)$ and it can be
expressed in terms of the scattering
matrix\cite{Gaspard-95,Smilansky-Ussishkin-95}.
The disorder averaging leads to
\begin{eqnarray}
    K^{\text{unit}}_\gamma(t) &\approx& K^{\text{unit}}_0(t) p(t) \;,
\label{eq:form.2}
\\
    K^{\text{unit}}_0(t) &=& \text{min}({t\Delta \over 2\pi}, 1)\;,
\label{eq:form.3}
\end{eqnarray}
where the superscript ``unit'' means that the form-factor was computed as if the
system is not symmetrical under the time reversal, and $p(t)$ is given by
Eq.~(\ref{eq:quant.4}). From the definition Eq.~(\ref{eq:form.1}) one can
see  that the form factor of the delay time $K_\gamma(t)$ approaches the form
factor of the density of states $K_0(t)$  when $\gamma$ goes to zero and this
is consistent with Eqs.~(\ref{eq:quant.4}) and (\ref{eq:form.2}). The decay
rate of the lowest diffusion mode, $\gamma\rightarrow0$,
when the opening of the cavity becomes smaller $L_1\rightarrow0$,  see
Eqs.~(\ref{eq:AppA.2.a}) and (\ref{eq:AppA.6}). In this case all particles
remains in the cavity forever and our solution gives $p(t)\rightarrow1$.

Our results are inconsistent with the random matrix
theory\cite{HDM-92,Lehmann-apr95,Fyodorov-may97,Savin-Sokolov-97}, which
predicts a power law decay of both $K_\gamma(t)$ and $p(t)$ if
the number of open channels is much smaller than the dimensionality of the 
Hamiltonian.  However, on the short time scale,  $t\Delta\ll1$, one has from 
Eq.~(\ref{eq:quant.4})
\begin{equation}
   -\log(p(t)) = \gamma t\left(1-{t\Delta \over 6\pi}\right) 
\label{eq:muda.1}
\end{equation}
and this is similar to the numeric results of Ref.\onlinecite{Casati-dec97}
and the random matrix theory calculations of Ref.\onlinecite{Frahm-dec97}.
The relation between $K_\gamma(t)$ and $p(t)$ similar to
Eq.~(\ref{eq:form.2}) appears in the random matrix
theory\cite{Savin-Sokolov-97} too.

Before going on to explain the averaging procedure let us give the semiclassical
path -- integral interpretation of the above result. Expansion of the Green
function in a sum over classical paths $j$ from the point $\vec r_1$ to
$\vec r_2$ is
\begin{equation}
    G^{R,A}_{E}(\vec r_1, \vec r_2) =
    \sum_{j} A_j \exp\{\pm i S_j(\vec r_1, \vec r_2, E) \}
\label{eq:semi.1}
\end{equation}
where $S_j(\vec r_1, \vec r_2, E)$ is the action along the path and
$A_j$ are some coefficients.\cite{Gutzwiller-book91} Therefore, the probability
to remain is the double sum over trajectories, which can be separated into
the diagonal and off-diagonal parts. Stationary phase integration over
$\omega$ in Eq.~(\ref{eq:quant.3}) gives
\begin{equation}
   p_{\text{diag}}(t) =
   \int_S d\vec r_1 d\vec r_2 \sum_{j} {A_j^2\over \nu S}
   \delta[t - T_j(\vec r_1, \vec r_2, E_F)] \;,
\label{eq:semi.2}
\end{equation}
where $T_j$ is the time which takes the particle with energy $E_F$ to go from
the point $\vec r_1$ to the point $\vec r_2$ along the path $j$.
Equation~(\ref{eq:semi.2}) is equivalent to Eq.~(\ref{eq:class.1}) and
therefore $p_{\text{diag}}(t)=e^{-\gamma t}$ is the classical probability to
remain. Therefore, the log-normal tail of the quantum probability to remain
represented in Eq.~(\ref{eq:quant.4})  is determined by the off-diagonal part of the sum
over trajectories in Eqs.~(\ref{eq:quant.3}) and (\ref{eq:semi.1}).

It is known that the integrals over coordinates in Eq.~(\ref{eq:semi.2}) can
be computed by using the skeleton of the periodic
orbits\cite{Cvitanovic-90,Cvitanovic-Eckardt-91}
\begin{equation}
   p_{\text{diag}}(t) =
   \sum_j T_j \sum_{r=1}^\infty {\delta(t - rT_j) \over |\det(I-M_j^r)|}
\label{eq:semi.3}
\end{equation}
where the index $j$ runs over all primitive periodic orbits inside the cavity
in configuration space and on the Fermi surface in phase space.
Each orbit has the period $T_j$ and the monodromy matrix $M_j$. From
Eq.~(\ref{eq:semi.3}) one concludes that the decay rate of
$p_{\text{diag}}(t)$ is equal to the difference between the Lyapunov exponent
and the Kolmogorov -- Sinai entropy and should be equal to the decay rate of
the lowest diffusion mode.\cite{Gaspard-96}

The semiclassical form of the Green functions Eq.~(\ref{eq:semi.1}) can be used
for computation of the form-factor Eq.~(\ref{eq:form.1}). The stationary
phase integration over $\vec r_1, \vec r_2, \omega$ gives\cite{Argaman-feb93}
\begin{equation}
    K^{\text{unit}}_{\gamma\,\text{diag}}(t)
    =  {t\Delta \over 2\pi}  p_{\text{diag}}(t) \;,
\label{eq:semi.4}
\end{equation}
which is in agreement with our result Eq.~(\ref{eq:form.2}) taken for short
times.

Let us assume that the leakage is so small, that all gradients are small on
the scale of the mean free path. The averaging of the product of two Green
functions from Eq.~(\ref{eq:quant.3})  is given by the functional integral
\begin{eqnarray}
    &&
    \left\langle
       G^R G^A
    \right\rangle
    = 2 (\pi\nu)^2 \int_{Q^2=1} Q^{12}_{\alpha\beta}(\vec r_1)
    k_{\beta\beta}
    Q^{21}_{\beta\alpha}(\vec r_2)
    e^{-F[Q]} DQ
\nonumber\\ &&
\label{eq:quant.5}
\\ &&
   F[Q] = {\pi \nu \over 8}\mbox{str}
   \int_S \biggl[ D(\nabla Q)^2
   + 2i(\omega + i\delta)\Lambda Q
   \biggr] d\vec r
\nonumber\\ &&
\label{eq:quant.6}
\\ &&
    D\tau\,\mbox{str}(\nabla Q)^2 \;\ll\; 1
\label{eq:quant.6cond}
\end{eqnarray}
where $D$ is the diffusion coefficient on the Fermi
surface, see Ref. \onlinecite{Efetov-book97}, page 60, $Q=\Lambda$ at the
contact, and $\vec n \vec \nabla Q = 0$ at the
walls.\cite{MK-oct94,Mirlin-oct95}

One can use the polar decomposition of $Q$ for calculation of the integral.
The probability to return should be independent of the symmetry of the system
and we will use the simplest ``unitary'' case. Then the action
Eq.~(\ref{eq:quant.6}) has to be multiplied by a factor of 2 and
\begin{eqnarray*}
   Q^{12} &=& iu\sin\hat\theta v^{-1}\;\;\;\;
   Q^{21} = -iv\sin\hat\theta u^{-1} \;\;\;\;
   \hat\theta =
   \left(\begin{array}{cc}
        \theta & 0\\ 0 & i\theta_1
    \end{array}\right)
\\
  u &=& \left(\begin{array}{cc}
      1 - 2\eta\eta^\ast & 2\eta \\
      - 2\eta^\ast & 1 + 2\eta\eta^\ast
  \end{array}\right)
   \left(\begin{array}{cc}
        e^{i\phi} & 0\\ 0 & e^{i\chi}
    \end{array}\right)
\\
  u^{-1} &=&
     \left(\begin{array}{cc}
        e^{-i\phi} & 0\\ 0 & e^{-i\chi}
    \end{array}\right)
    \left(\begin{array}{cc}
      1 - 2\eta\eta^\ast & -2\eta \\
      2\eta^\ast & 1 + 2\eta\eta^\ast
  \end{array}\right)
\\
  v &=& \left(\begin{array}{cc}
      1 + 2\kappa\kappa^\ast & 2i\kappa \\
      - 2i\kappa^\ast & 1 - 2\kappa\kappa^\ast
  \end{array}\right)\;,
\\
  v^{-1} &=& \left(\begin{array}{cc}
      1 + 2\kappa\kappa^\ast & -2 i \kappa \\
      2 i \kappa^\ast & 1 - 2\kappa\kappa^\ast
  \end{array}\right)
\end{eqnarray*}
where $\theta$, $\theta_1$, $\phi$, $\chi$ are commuting variables and
$\kappa$, $\kappa^\ast$, $\eta$, $\eta^\ast$ are anticommuting variables
which parameterize the matrix $Q$. Let us write explicitly the elements of
the matrix $Q$ which appear in the integrand in Eq.~(\ref{eq:quant.5})
\begin{eqnarray*}
&&
    Q^{12} =
    i
\\ &&\times
    \left(\begin{array}{c}
    [1 - 2\eta\eta^\ast][1 + 2\kappa\kappa^\ast] e^{i\phi}\sin \theta
    -4  \eta\kappa^\ast e^{i\chi}\sinh\theta_1
    \\
    - 2\eta^\ast[1 + 2\kappa\kappa^\ast] e^{i\phi}\sin \theta
    - 2[1 + 2\eta\eta^\ast] \kappa^\ast
    e^{i\chi}\sinh\theta_1\end{array}\right.
\\  &&
    \left.\begin{array}{c}
    -2 i [1 - 2\eta\eta^\ast]  \kappa e^{i\phi}\sin \theta +
    2i\eta[1 - 2\kappa\kappa^\ast] e^{i\chi}\sinh\theta_1
    \\ 4 i\eta^\ast  \kappa e^{i\phi}\sin \theta
    + i[1 + 2\eta\eta^\ast][1 - 2\kappa\kappa^\ast]
    e^{i\chi}\sinh\theta_1
    \end{array}\right)
\\ && Q^{21} = -i
\\ &&\times
    \left(\begin{array}{c}
    [1 + 2\kappa\kappa^\ast][1 - 2\eta\eta^\ast]e^{-i\phi}\sin\theta
    -4 \kappa \eta^\ast e^{-i\chi}\sinh\theta_1 \\
    - 2i\kappa^\ast[1 - 2\eta\eta^\ast]e^{-i\phi}\sin\theta
    + 2i[1 - 2\kappa\kappa^\ast]\eta^\ast
    e^{-i\chi}\sinh\theta_1\end{array}\right.
\\  &&
    \left.\begin{array}{c}
    -2[1 + 2\kappa\kappa^\ast]\eta e^{-i\phi}\sin\theta
    -2\kappa[1 + 2\eta\eta^\ast]e^{-i\chi}\sinh\theta_1 \\
    +4i\kappa^\ast\eta e^{-i\phi}\sin\theta
    +i[1 - 2\kappa\kappa^\ast][1 + 2\eta\eta^\ast]e^{-i\chi}\sinh\theta_1
    \end{array}\right)
\end{eqnarray*}

Our purpose is to reduce the functional integral over the super-matrix $Q$ to 
the
conventional integral. The minimum of the action is reached when $u$ and $v$
are independent of coordinates\cite{MK-oct94} and the action becomes
\begin{eqnarray}
   F &=& {\pi \nu \over 2}
   \int_S \biggl\{
   D(\nabla \theta)^2 + D(\nabla \theta_1)^2
\nonumber\\
   &+& 2(i\omega-0) (\cos\theta - \cosh\theta_1)
   \biggr\} d\vec r\;,
\label{eq:quant.7}
\end{eqnarray}
where $\hat \theta = 0$ at the contact and $\vec n \vec \nabla \hat \theta =0$
at the walls. The same boundary conditions were
applied to Eq.~(\ref{eq:class.2}) and therefore we can use the diffusion
modes for computing the functional integral over $\hat \theta(r)$.
Due to the condition Eq.~(\ref{eq:quant.4a}) only the lowest diffusion mode
contributes and the functional integral becomes a conventional integral over the
amplitude of this mode $\hat\Theta$. The lowest diffusion mode is almost
uniform, see Appendix~\ref{sec:App.A}, and therefore
\begin{eqnarray*}
 {1\over S}     \int_S \theta^2 d\vec r &=&\Theta^2 \;,\\
 {1\over S}     \int_S \theta_1^2 d\vec r &=&\Theta_1^2 \;,\\
 {1\over S}  \int_S D(\nabla\theta)^2 d\vec  r &=&\gamma\Theta^2 \;,\\
 {1\over S}  \int_S \cosh\theta d\vec r &=&\cosh\Theta \;,\\
 {1\over S}  \int_S \sinh\theta d\vec r &=&\sinh\Theta \;,
\end{eqnarray*}
and so on for other functions of $\theta$ and $\theta_1$. Particularly
the pre-exponential factor in the expression for the return probability
becomes
\begin{equation}
  Q^{12}_{21} Q^{21}_{12}
  = -8\eta\eta^\ast\kappa\kappa^\ast(\sin^2\Theta+\sinh^2\Theta_1)
\label{eq:quant.9}
\end{equation}
We computed this factor explicitly from the expressions for $Q^{12}$ and $Q^{21}$.
This expression was computed in Appendix 3 of Ref.\onlinecite{Efetov-book97}
but the numeric coefficients are different.

The model reduces to the conventional integral
\begin{eqnarray}
       &&
    p(t) =  -2(\pi\nu)^2\int {d\omega\over 2\pi}\,e^{-i\omega t}
    \int
    \underbrace{8\eta\eta^\ast\kappa\kappa^\ast
    (\sin^2\Theta+\sinh^2\Theta_1)}_{=Q^{12}_{21} Q^{21}_{12}}
\nonumber\\ &&\times \,e^{-F}\,
   \underbrace{2^{-8}(\cosh\Theta_1-\cos\Theta)^{-2}}_{=J_1}
   \underbrace{d\eta d\eta^\ast d\kappa^\ast d\kappa}_{=dR_1}
\nonumber\\ &&\times \,\underbrace{(2/\pi)^2\sin\Theta \sinh\Theta_1}_{=J_2}
    \underbrace{d\phi d\chi}_{=dR_2} d\Theta d\Theta_1
    = (\pi\nu)^2\int {d\omega\over 2\pi}\,e^{-i\omega t}
\nonumber\\ & &
    \times\int \,e^{-F}\,
    { \cosh\Theta_1+\cos\Theta\over \cosh\Theta_1-\cos\Theta}
   \sin\Theta \sinh\Theta_1 d\Theta d\Theta_1
\label{eq:quant.10}
\\ && F = {\pi \over \Delta}
   \biggl\{
   \gamma{ \Theta^2+\Theta_1^2\over 2 }
   + (i\omega-0) (\cos\Theta -\cosh\Theta_1 )
   \biggr\}\;,
\label{eq:quant.11}
\end{eqnarray}
where we have kept Efetov's notations  for differentials and Jacobians,
see Ref.~\onlinecite{Efetov-book97}, page 107.

The standard change of variables $\lambda = \cos \Theta$ and
$\lambda_1  = \cosh \Theta_1 $ leads to the
relatively simple expression for the probability to  remain
\begin{eqnarray}
   p(t) &=& {1 \over 2 }
   \int_{\text{max}(-1, 1-t\Delta/\pi)}^1 d\lambda
   \;{ t\Delta/\pi +2\lambda\over t\Delta/\pi}
\nonumber\\  &\times &
    e^{-{\pi \gamma\over 2\Delta}
    ( \arccos^2(\lambda)+\text{arccosh}^2(t\Delta/\pi +\lambda) )}\;.
\label{eq:quant.12}
\end{eqnarray}
For both short times $t\ll \pi/\Delta$ and long times $t\gg \pi/\Delta$
we can put $\lambda=1$ in the exponent on the right hand side of
Eq.~(\ref{eq:quant.12}). Then the integral can be computed exactly and we
arrive at our interpolation formula Eq.~(\ref{eq:quant.4}). The computation
of the form-factor is similar,
\begin{eqnarray}
   K_\gamma^{\text{unit}}(t)
   &=& {1 \over 2 }
   \int_{\text{max}(-1, 1-t\Delta/\pi)}^1 d\lambda
\nonumber\\  &\times &
    e^{-{\pi \gamma\over 2\Delta}
    ( \arccos^2(\lambda)+\text{arccosh}^2(t\Delta/\pi +\lambda) )}\;.
\label{eq:quant.14}
\end{eqnarray}
and Eq.~(\ref{eq:form.2}) matches both short and long time asymptotes
of Eq.~(\ref{eq:quant.14}). Both results Eqs.~(\ref{eq:quant.12}) and
(\ref{eq:quant.14}) where derived under the condition Eq.~(\ref{eq:quant.6cond})
which becomes Eq.~(\ref{eq:quant.4c}) in the one-mode approximation.

The functional integral over $\hat \theta(\vec r)$ in the theory with action
Eq.~(\ref{eq:quant.7}) can be computed by the optimal fluctuation
method\cite{MK-oct94,Mirlin-oct95,Falko-Efetov-95}. The calculations are
straightforward and give the same results, see Appendix~\ref{sec:App.B},
where we have checked the three dimensional case. In the same way one can also
check the two dimensional result.

In the present work we have assumed that the opening or contact is small. If
it is not the case for some system the result for the probability to remain
may be different. In such a system one should consider the optimal
fluctuations of the random potential\cite{Smolyarenko-Altshuler-97}. 
In our geometry this method can be used for very long times, which broke 
the condition Eq.~(\ref{eq:quant.4c}).

In summary, we have obtained the arccosh-normal distribution for the
probability to remain in a disordered cavity. We made use of the
one-mode approximation, which is similar to both the
zero-dimensional $\sigma$-model and the optimal fluctuation method. Our result
is not consistent with the random matrix theory prediction.

\acknowledgments
It is my pleasure to thank prof. Uzy Smilansky for important remarks and
discussions.
This work was supported by Israel Science Foundation and the Minerva Center
for Nonlinear Physics of Complex systems.

\appendix

\begin{figure}
\unitlength=1mm
\linethickness{0.4pt}
\begin{picture}(70.00,60.00)
\put(10.00,10.00){\framebox(60.00,50.00)[cc]{}}
\bezier{340}(70.00,40.00)(15.00,40.00)(11.00,10.00)
\put(40.00,2.00){\makebox(0,0)[cb]{$r$}}
\put(70.00,8.00){\makebox(0,0)[ct]{$L_2$}}
\put(11.00,8.00){\makebox(0,0)[ct]{$L_1$}}
\put(3.00,35.00){\makebox(0,0)[cc]{$\varrho(r)$}}
\end{picture}
   \caption{Radial dependence of the diffusion mode for 
   $\ell\ll L_1\ll L_2$.}
\label{fig:profile}
\end{figure}
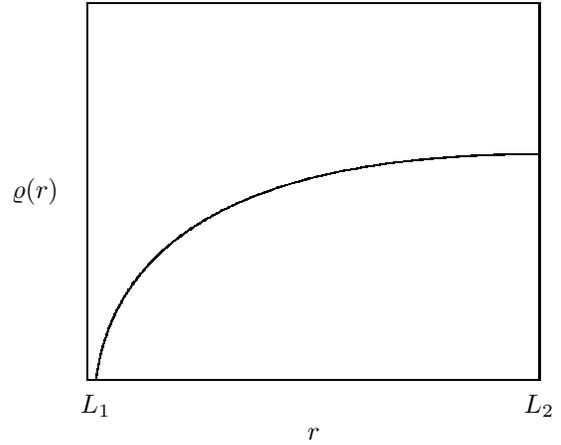

\section{ Computation of the diffusion mode. }
\label{sec:App.A}

Let us compute the diffusion mode for the circular cavity with the circular
contact in the middle, as shown in Fig.~\ref{fig:circular}.
The lowest mode is independent of the polar angle and the radial dependence
is expressed in terms of the Bessel functions
\begin{eqnarray}
  \varrho(r) &\propto&
  \mbox{Im}\Bigl[
  H_0^{(1)}(r\sqrt{\gamma\over D})
  H_1^{(2)}(L_2\sqrt{\gamma\over D})
  \Bigr]
\label{eq:AppA.1}
\\
  0 &=&
  \mbox{Im}\Bigl[
  H_0^{(1)}(L_1\sqrt{\gamma\over D})
  H_1^{(2)}(L_2\sqrt{\gamma\over D})
  \Bigr]
\label{eq:AppA.2}
\end{eqnarray}
where $L_2\sqrt{\gamma\over D}$ has to be less that the first root of $J_1(x)$
and this is possible for roughly $L_1/L_2 < 1/2$.
The solution for $L_1\ll L_2$ is shown in Fig.~\ref{fig:profile} and in this
case
\begin{equation}
   \gamma = {D\over L_2^2 \log{L_2\over L_1}}\;.
\label{eq:AppA.2.a}
\end{equation}
We see that the mode is non-uniform for roughly $L_1<r<L_2/2$. Therefore
the presence of the contact strongly affects  the diffusion mode in a quarter
of the cavity area.

In the case of the three dimensional sphere it might be difficult to maintain
the 
contact in the middle, but it is still interesting to compute the lowest
diffusion mode. In three dimensions the lowest mode is independent of polar
and azimuthal angles and depends only on the distance $r$ from the center.
The lowest mode in three dimensions is
\begin{eqnarray}
   \varrho(r)   &\propto&
   {1\over r} \sin\sqrt{3L_1{(r-L_1)^2\over L_2^3}}
\label{eq:AppA.3}
\\
   L_2\sqrt{\gamma\over D} &=& \tan \sqrt{\gamma{(L_2-L_1)^2\over D}}
   \;,
\label{eq:AppA.4}
\end{eqnarray}
where $L_2$ is the radius of the cavity and $L_1$ is the radius of the
contact. For the small values of the ratio $L_1/L_2$ this mode is also
uniform in the most of the volume. In this case the three dimensional mode
becomes
\begin{eqnarray}
   \varrho(r)&\propto&
   {r-L_1\over r}  - {L_1 \over 2 r}\,\left({r-L_1\over L_2}\right)^2
\label{eq:AppA.5}
\\
   \gamma &=& 3L_1 D/L_2^3
\label{eq:AppA.6}
\end{eqnarray}
and its shape is similar to that which is shown in Fig.~\ref{fig:profile}.

\section{ The optimal fluctuation method in three dimensions. }
\label{sec:App.B}

According to the optimal fluctuation method the probability to remain is
given by the following system of equations\cite{MK-oct94}
\begin{eqnarray}
   p(t) &\propto& e^{-{\pi \nu\over 2}D\int_S (\nabla \theta)^2  d\vec r }\;,
\label{eq:AppB.1}
\\
   t &=& \pi \nu \int_S \cosh(\theta) d\vec r \;,
\label{eq:AppB.2}
\\
   D\nabla^2 \theta &=& - i\omega \sinh(\theta)\;,
\label{eq:AppB.3}
\end{eqnarray}
and the boundary conditions for the last equation were written after
Eq.~(\ref{eq:quant.7}).  This very problem was solved by Falko and
Efetov\cite{Falko-Efetov-95}, Sec.~VI and VII. In three dimensions one has
\begin{eqnarray}
   \theta &=& A(1-{L_1\over r})  \;,
\label{eq:AppB.4}
\\
   A/e^A &=& {i\omega L_2^3 \over 3L_1 D} \ll 1 \;,
\label{eq:AppB.5}
\\   t &=& {\pi \over \Delta}\;e^A
\label{eq:AppB.6}
\\
   p(t) &\propto&
     e^{-{3\pi D\over 2\Delta}\, {A^2L_1\over L_2^3} }
     =
     e^{-{\pi \gamma\over 2\Delta} \log^2{t\Delta\over \pi}}
\label{eq:AppB.7}
\end{eqnarray}
where $\gamma$ is given by Eq.~(\ref{eq:AppA.6}). This result is precisely
equal to our one mode result Eq.~(\ref{eq:quant.4}) in the long time limit.


\end{multicols}
\end{document}